\documentclass[aps,prb,twocolumn,citeautoscript,superscriptaddress,floatfix]{revtex4-1}

\usepackage{graphics}
\usepackage{graphicx}
\usepackage{amsmath}
\usepackage{amssymb}
\usepackage{times}

\newcommand{\vect}[1]{{\mathbf{#1}}}
\newcommand{\argrs}{\left(\vect{r},s\right)}

\begin{document}

\title{Coherent States Formulation of Polymer Field Theory}
\author{Xingkun Man}
\affiliation{Department of Chemical Engineering, University of California, Santa Barbara, California 93106, USA}
\affiliation{Materials Research Laboratory, University of California, Santa Barbara, California 93106, USA}
\author{Kris~T.~Delaney}
\affiliation{Materials Research Laboratory, University of California, Santa Barbara, California 93106, USA}
\author{Michael~C.~Villet}
\affiliation{Department of Chemical Engineering, University of California, Santa Barbara, California 93106, USA}
\affiliation{Materials Research Laboratory, University of California, Santa Barbara, California 93106, USA}
\author{Henri Orland}
\affiliation{Institut de Physique Th\'eorique, CE-Saclay, CEA, F-91191 Gif-sur-Yvette Cedex, France}
\author{Glenn~H.~Fredrickson}
\affiliation{Department of Chemical Engineering, University of California, Santa Barbara, California 93106, USA}
\affiliation{Materials Research Laboratory, University of California, Santa Barbara, California 93106, USA}
\affiliation{Materials Department, University of California, Santa Barbara, California 93106, USA}

\begin{abstract}
We introduce a stable and efficient complex Langevin (CL) scheme to enable the first direct numerical simulations of the
coherent-states (CS) formulation of polymer field theory.
In contrast with Edwards' well-known auxiliary-field (AF) framework, the CS formulation does not contain an
embedded nonlinear, non-local, implicit functional of the auxiliary fields, and
the action of the field theory has a fully explicit, semi-local and finite-order polynomial character.
In the context of a polymer solution model, we demonstrate that the new CS-CL dynamical scheme for sampling fluctuations
in the space of coherent states yields results in good
agreement with now-standard AF-CL simulations.
The formalism is potentially applicable to a broad range of polymer architectures and may facilitate systematic generation of trial
actions for use in coarse-graining and numerical renormalization-group studies.

\end{abstract}

\maketitle

\section{Introduction}
\label{sec:intro}

The standard auxiliary-field (AF) polymer field theory introduced by Edwards in the 1960s\cite{Edwards1965}
has proven to be a powerful tool for investigating the properties of mesoscopic models of inhomogeneous
polymers at equilibrium and has been well developed on both analytical and numerical fronts over the past five decades.
Applying the mean-field approximation to Edwards' framework results in self-consistent field theory (SCFT),
which has been the basis for exploring the self-assembling structure of a wide range of important heterogeneous
polymer systems\cite{Matsen1994,Glenn2006,Zuo2008}.
Moreover, field-theoretic simulations (FTS), which move beyond the confines of the mean-field
approximation, offer a scalable and general computational methodology for studying fluctuation
phenomena in Edwards' AF-type models, with numerical methods for circumventing the sign problem
based on complex Langevin (CL) sampling\cite{Ganesan2001,Glenn2006,Katz2003,Popov2007,Lennon2008a}.

The standard approach to formulating the Edwards AF framework involves the use of
Hubbard-Stratonovich transformations to decouple segmental interactions through the introduction
of one or more spatially-varying auxiliary potential fields\cite{Edwards1966,Helfand1975,Hong1981}.
The effective Hamiltonian resulting from this procedure is a nonlinear functional of
the auxiliary fields, with the non-linearity embedded within a single-chain partition function that
enumerates the conformational states of decoupled polymer chains subject to an external field.
The non-linear and non-local character of such Hamiltonians cause difficulties for advanced applications of
polymer field theory, such as the systematic coarse graining (CG) and numerical renormalization-group (RG)
studies of a wide range of polymeric fluids, and investigations of rigid or semi-flexible
liquid-crystalline polymers\cite{Villet2010, VilletThesis}.

In 1970, Edwards and Freed introduced\cite{EdwardsFreed1970} an alternative formulation of polymer field
theory based on the second-quantized description of the quantum many-body problem, and adapted it to
analytical studies of the statistical thermodynamics of an assembly of crosslinked polymers.
However, the formalism has been largely neglected since that time, has not been explored in the context of
conventional linear (non-crosslinked) polymer solutions and melts and, to our knowledge, has never been
applied as a basis for numerical simulations.
The distinguishing feature of this formalism is the absence of auxiliary external fields; the only fields
entering the effective Hamiltonian are propagator-like ``coherent states'' (CS).

The CS formalism presents several attractive features.
First, the Hamiltonian of the CS formalism is an explicit fourth-order functional of the fields,
whereas the Edwards Hamiltonian is a nonlinear functional containing terms of all powers in the
auxiliary fields. Second, while the Edwards Hamiltonian is non-local due to the single-chain partition
function, the CS Hamiltonian has a simple semi-local character (i.e., gradient terms) that is explicit.
We expect that these features can be exploited to improve the numerical efficiency of both mean-field and
fully fluctuating simulations for a broad range of polymer field theories.
Furthermore, the simple structure of the CS Hamiltonian should facilitate multi-scale simulations,
in particular, through easier identification of suitable trial coarse-grained (CG) action functionals
for variational force-matching\cite{Villet2010, VilletThesis}.
Finally, one might hope that these characteristics of the CS theory would permit convenient extension
to branched and networked polymers, rigid or semi-flexible chains, and polyelectrolytes,
all of which present challenges for numerical simulations in the conventional AF framework.

In this paper, we revisit the Edwards-Freed CS formalism in the context of
a standard polymer solution model and introduce a stable and accurate stochastic dynamics scheme
for conducting fully fluctuating numerical simulations.
\section{Theoretical Framework}
\label{sec:theory}

\subsection{Transformation to the CS formulation: Grand Canonical Ensemble of a Polymer Solution Model}
\label{sec:cstheory}
The CS description of a polymer field theory can be derived directly from the standard
auxiliary-field formalism.
We use the Edwards model of a grand-canonical ensemble of homopolymers in an implicit solvent
as a platform to develop the theory and algorithms necessary to conduct simulations in the CS framework.
The auxiliary-field grand partition function for this model can be written as\cite{Glenn2006}
\begin{eqnarray}\label{eqn:GCEAF}
  &\Xi&\left(  z^\prime  ,V,T\right) = \int \mathcal{D}w\,
\sum^{\infty}_{n=0} \frac{\left(z^\prime VQ\left[iw\right]\right)^n}{n!}\times\nonumber\\
& &\exp\left(-\frac{1}{2}\int d\vect{r}\,\int d\vect{r}^{\prime}\,
 w\left(\vect{r}\right)u^{-1}\left(\left|\vect{r}-\vect{r}^{\prime}\right|\right)w\left(\vect{r}^{\prime}\right) \right),
\end{eqnarray}
where $Q\left[iw\right]$ is the single-chain partition function in the presence of the purely imaginary
auxiliary field $iw\left(\vect{r}\right)$, $u\left(r\right)=u_0\delta\left(r\right)$ is the on-contact excluded-volume pair
interaction between monomers, with $u_0$ in units of $k_B T$, and $z^\prime$ is a polymer chain activity referenced
to an ideal gas of polymers with the same chain statistics.
The partition function $Q\left[iw\right]$ can be rewritten as a linear functional of a chain-end
Green function, $VQ\left[iw\right] = \int d\vect{r}
\int d\vect{r}^\prime \, G\left(\vect{r},N \mid \vect{r}',0 ; \left[iw\right]\right)$,
where $\vect{r}$ and $\vect{r}^{\prime}$
represent the spatial coordinates of the two polymer ends and $N$ is the polymerization degree.
By definition, $G\left(\vect{r},s\mid \vect{r}^\prime,s^\prime;\left[iw\right]\right)$ is the
inverse of the operator
\begin{eqnarray}\label{eqn:mdeoperator}
  \mathcal{L}=\partial_s-\frac{b^2}{6}\nabla^2 + iw\left(\vect{r}\right),
\end{eqnarray}
appearing in the Fokker-Planck equation for a polymer chain propagator
(segmental probability distribution) with continuous Gaussian stretching statistics\cite{Glenn2006}.
In order to represent the Green function, we introduce the following generating functional
\begin{widetext}
\begin{equation}
  Z\left[h,\hat{h}\right] = \int\mathcal{D}\varphi\int\mathcal{D}\hat{\varphi}\,
  e^{-\mathcal{S}\left[\varphi,\hat{\varphi}\right]+
  \int d\vect{r}\int ds\, h\left(\vect{r},s\right)\varphi\left(\vect{r},s\right) +
 i\int d\vect{r}\int ds\, \hat{h}\left(\vect{r},s\right)\hat{\varphi}\left(\vect{r},s\right)}
\end{equation}
\end{widetext}
where the free-field action is
$\mathcal{S}\left[\varphi,\hat{\varphi}\right] =
-i\int_0^N ds\int d\vect{r}\, \hat{\varphi}\left(\vect{r},s\right)\mathcal{L}\varphi\left(\vect{r},s\right)$,
and the functional integrals $\mathcal{D}\varphi$ and $\mathcal{D}\hat{\varphi}$ are understood to be over all
possible domain-supported real-valued fields.
This generating functional is familiar both in the
coherent states representation of quantum field theory\cite{NegeleOrland}
and the path integral formulation of Martin-Siggia-Rose stochastic
classical dynamics\cite{MSR,Jensen1981}.
We employ a causal chain contour discretization in Jensen's framework, which
produces a field-independent Jacobian in the functional integral over $\varphi$.
It is straightforward to show that
\begin{eqnarray}
  \label{eqn:Gfn}
  -\left.\frac{\delta^2 \ln Z\left[h,\hat{h}\right]}{\delta h\left(\vect{r},s\right)\delta
  \hat{h}\left(\vect{r}^\prime,s^\prime\right)}\right|_{h=\hat{h}=0} & = &
  G\left(\vect{r},s\mid\vect{r}^\prime,s^\prime; \left[iw\right]\right)\\
  &=& -i\left<\varphi\left(\vect{r},s\right)\hat{\varphi}\left(\vect{r}^\prime,s^\prime\right)\right>_0,\nonumber
\end{eqnarray}
where
\begin{equation}
  \left<O\left[\varphi,\hat{\varphi}\right] \right>_0
= \frac{1}{Z\left[0,0\right]}\int \mathcal{D}\varphi\int\mathcal{D}\hat{\varphi}\,
O\left[\varphi,\hat{\varphi}\right] e^{-\mathcal{S}\left[\varphi,\hat{\varphi}\right]},
\end{equation}
and the first moments of $\varphi$ and $\hat{\varphi}$ vanish under zero $h$, $\hat{h}$.
The Green function in Eqn.~\ref{eqn:Gfn} is causal and vanishes for $s\leq s^\prime$,
and all other second moments of the $\varphi$ and $\hat{\varphi}$ fields vanish
identically.
Using these relationships and Wick's theorem, one can enumerate the contractions of
all powers of the integrated chain-end Green function to yield
\begin{widetext}
\begin{equation}\label{eqn:GtoCS}
\sum^{\infty}_{n=0}\frac{\left[z^{\prime}\int d\vect{r}\int d\vect{r}^{\prime}\,
G(\vect{r},N\mid\vect{r}^{\prime},0;\left[iw\right])\right]^n}{n!}
= \frac{\int \mathcal{D}\varphi\int \mathcal{D}\hat{\varphi} \,
e^{-\mathcal{S}\left[\varphi,\hat{\varphi}\right]+\sqrt{z^{\prime}}\int d\vect{r}
\left[-i\hat{\varphi}(\vect{r},0)+\varphi(\vect{r},N)\right]}}{\int \mathcal{D}\varphi \int \mathcal{D}\hat{\varphi}\, e^{-\mathcal{S}\left[\varphi,\hat{\varphi}\right]}},
\end{equation}
\end{widetext}
The denominator in Eqn.~\ref{eqn:GtoCS} can be ignored: provided an appropriate causally respectful discretization scheme is
used for the contour variable $s$, such as the one provided in Appendix \ref{sec:discrete}, the denominator is a
constant that is independent of the field configuration
$iw$ and therefore has no thermodynamic consequences.
A numerical demonstration of the causal nature of our discretization scheme is
provided in Sec.~\ref{sec:Gfn}.
%
Inserting Eqn.~\ref{eqn:GtoCS} into Eqn.~\ref{eqn:GCEAF} and completing the Gaussian functional
integral over the now-explicit $iw(\vect{r})$ auxiliary field leads to
\begin{equation}\label{eqn:GCECS}
  \Xi\left(z^\prime ,V,T\right)=\Xi_0\int \mathcal{D}\varphi\int \mathcal{D}\hat{\varphi}\, \exp\left(-H\left[\hat{\varphi},\varphi\right]\right),
\end{equation}
with effective Hamiltonian (action) functional
\begin{eqnarray}\label{eqn:CSaction_dimful}
H\left[\hat{\varphi},\varphi\right]&=&-i\int^{N}_{0} ds\int d\vect{r} \,\hat{\varphi}\left(\vect{r},s\right)
     \left(\partial_s-\frac{b^2}{6}\nabla^2\right)\varphi\left(\vect{r},s\right)\nonumber\\
& +&\frac{1}{2}\int d\vect{r}\int d\vect{r}^{\prime}\,\hat{\rho}\left(\vect{r}\right)u\left(\left|\vect{r}-\vect{r}^{\prime}\right|\right)\hat{\rho}\left(\vect{r}^{\prime}\right)\nonumber\\
& -&\sqrt{z^{\prime}}\int d\vect{r}\,\left[-i\hat{\varphi}\left(\vect{r},0\right)+\varphi\left(\vect{r},N\right)\right],
\end{eqnarray}
where $\hat{\rho}$ is a monomer density operator given by
\begin{equation}\label{f7}
\hat{\rho}\left(\vect{r}\right)=-i\int^{N}_{0}ds\,\hat{\varphi}\left(\vect{r},s\right)\varphi\left(\vect{r},s\right),
\end{equation}
and $\Xi_0$ is a constant from the denominator term.
The terms in $H\left[\hat{\varphi},\varphi\right]$ have an intuitive interpretation, with the first describing single-chain
statistics of non-interacting polymers, the second term their pairwise excluded-volume interactions, and the third a
source injection at the two chain ends necessary to generate the grand canonical ensemble of polymers.

We note that the original Edwards formulation of this model employed delta function excluded-volume
interactions for $u\left(r\right)$.
This choice, in which $u$ is not finite on contact, leads to an ultraviolet (UV) divergent theory,
in which averages of field-theoretic operators do not have a finite continuum limit\cite{Wang2010,Villet}.
To regularize the theory\cite{VilletThesis}, we instead adopt the repulsive Gaussian potential
\begin{eqnarray}\label{f8}
u\left(r\right) = \frac{u_0}{\left(4\pi a^2\right)^{3/2}} \exp\left( - \frac{r^2}{4 a^2}\right),
\end{eqnarray}
with strength $u_0$ and range parameter $a$.

For numerical convenience, we rescale $s\in\left[0,1\right]$, express all spatial
lengths in units of the free-polymer radius of gyration $R_g=b\sqrt{N/6}$,
and absorb a factor of $R_g^{3/2}$ into both fields, resulting in the rescaled Hamiltonian
\begin{eqnarray}\label{eqn:CSaction_dimless}
H\left[\hat{\varphi},\varphi\right]&=&-i\int^{1}_{0} ds\int d\vect{r} \, \hat{\varphi}\left(\vect{r},s\right)
  \left(\partial_s-\nabla^2\right)\varphi\left(\vect{r},s\right)\nonumber\\
  & &+\frac{B}{2}\int d\vect{r}\int d\vect{r}^{\prime}\,\hat{\rho}(\vect{r})
  \Gamma \left(|\vect{r}-\vect{r}^{\prime} | \right)
  \hat{\rho}\left(\vect{r}^{\prime}\right)\nonumber\\
& &-\sqrt{z}\int d\vect{r}\,\left[-i\hat{\varphi}\left(\vect{r},0\right)+\varphi\left(\vect{r},1 \right)\right],
\end{eqnarray}
where $B=u_0N^2/R^{3}_g$ is a dimensionless excluded-volume parameter,
$z=z^{\prime}R^{3}_g$ is a dimensionless chain activity,
$\Gamma\left(r\right) =
\exp\left(-r^2/\left(4\bar{a}^2\right)\right)/\left(4\pi\bar{a}^2\right)^{3/2} $
is a reduced potential function with range parameter $\bar{a} = a/R_g$,
and the dimensionless polymer chain density operator is
\begin{eqnarray}\label{f10}
\hat{\rho}(\vect{r})=-i\int^{1}_{0}ds\,\hat{\varphi}\left(\vect{r},s\right)\varphi\left(\vect{r},s\right).
\end{eqnarray}
%
\begin{figure*}[h!t]
\begin{center}
{\includegraphics[bb=0 0 350 239, scale=1.2,draft=false]{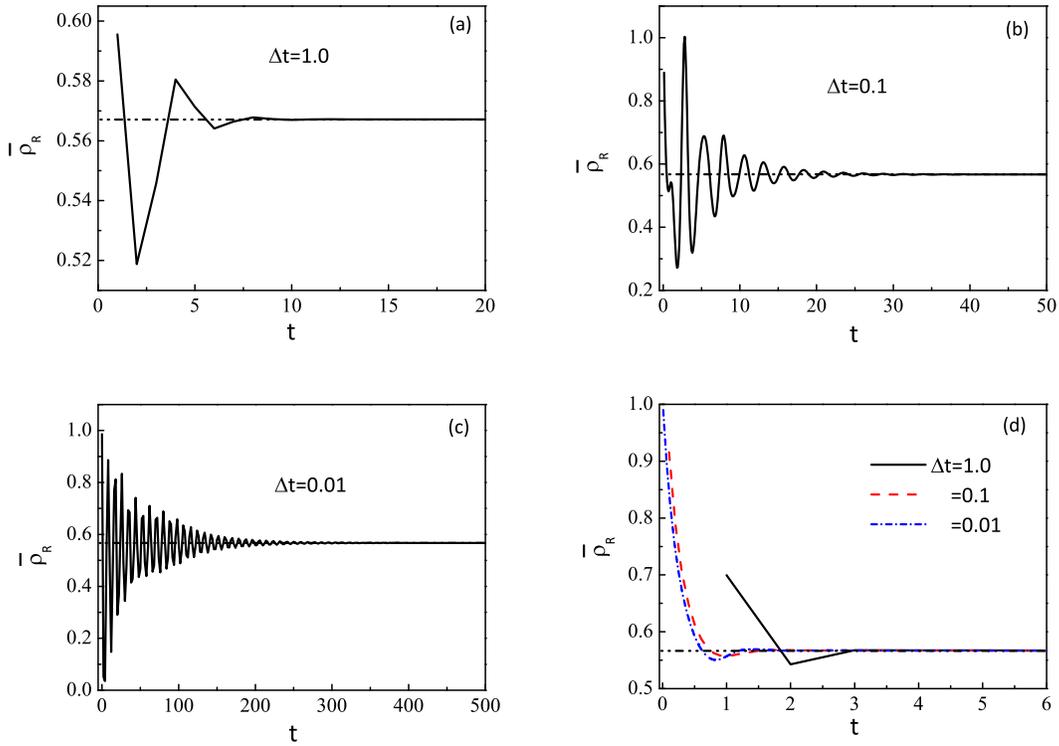}}
\caption{Trajectories for numerical relaxation of the spatially averaged chain
density, $\bar{\rho}_R$, to the mean-field value with various discrete time steps for both
the original relaxation scheme, (a), (b) and (c), and the modified scheme, (d). The
relaxation trajectories shown in (d) have no strong oscillations after inserting
a decoupling mobility matrix into the relaxation equations.
The dash-double-dot line is the exact mean-field value.
All calculations were conducted for $B=1$, $\Delta x=\bar{a}=0.1 R_g$, $L=0.8 R_g$,
and $\Delta s=0.01$ under periodic boundary conditions.}\label{fig:SCFT}
\end{center}
\end{figure*}
%
\subsection{Analytic Structure of the Saddle Point}
\label{sec:saddlept}
%
The saddle-point solution of the CS equations is defined in the usual way, by positing that the
partition function is dominated by a single extremal field conguration
\begin{equation}
  \Xi\left(z^\prime,V,T\right) \approx \exp\left(-H\left[\varphi^\star,\hat{\varphi}^\star\right]\right),
\end{equation}
where
\begin{equation}
  \left.\frac{\delta H\left[\varphi,\hat{\varphi}\right]}{\delta \varphi\argrs}\right|_{\varphi=\varphi^\star} = \left.\frac{\delta H\left[\varphi,\hat{\varphi}\right]}{\delta \hat{\varphi}\left(\vect{r},s\right)}\right|_{\hat{\varphi}=\hat{\varphi}^\star} = 0.
\end{equation}
For the CS field theory with Hamiltonian defined in Eqn.\ \ref{eqn:CSaction_dimless},
it is evident from the complex character of $H$ that
the physically relevant saddle-point solution must lie off the
real-valued integration domain for thermodynamic observables
to be real-valued.
We must therefore promote $\varphi, \hat{\varphi} \in \mathbb{C}$ and analytically continue $H$
in order to search for valid saddle-point solutions.

In the present case, it is sufficient to restrict $\varphi \in \mathbb{R}$ and $\hat{\varphi} \in -i\mathbb{R}$.
By choosing $\tilde{\varphi} = -i \hat{\varphi}$, a Wick rotation, one can search for a saddle point field configuration
with $\varphi, \tilde{\varphi} \in \mathbb{R}$.
Moreover, for this model, the $H$ functional is actually \emph{convex} for fields restricted to this domain,
rather than adopting the usual mixed saddle-point character often found in AF theories.
This feature, which we shall not explore in further detail in the present paper, may open the
possibility to employ more advanced and efficient SCFT optimization algorithms, such as conjugate gradients or,
by exploiting the explicit nature of the action with straightforward second functional derivatives, quasi-Newton minimizers.

We note that the model presented herein can be trivially extended to a
grand-canonical multi-species homopolymer blend.
In that case, with the same Wick rotation introduced above, the Hamiltonian functional is
\begin{eqnarray}\label{f11}
  H\left[\left\{\varphi_i\right\},\left\{\tilde{\varphi}_i\right\}\right]
  &=&\sum_i \int^{\alpha_i}_{0}ds\int d\vect{r} \, \tilde{\varphi}_i\argrs
  \left(\partial_s-\nabla^2\right)\varphi_i\argrs\nonumber\\
  & &+\sum_{i,j}\frac{B_{ij}}{2}\int d\vect{r}\int d\vect{r}^{\prime}\,\hat{\rho}_i(\vect{r})
  \Gamma \left(|\vect{r}-\vect{r}^{\prime} | \right)
  \hat{\rho}_j\left(\vect{r}^{\prime}\right)\nonumber\\
  & &-\sum_i\sqrt{z_i}\int d\vect{r}\,\left[\tilde{\varphi}_i\left(\vect{r},0\right)+\varphi_i\left(\vect{r},1 \right)\right],
\end{eqnarray}
where $B_{ij}$ is a matrix of excluded-volume interaction parameters between species $i$ and $j$, and $\alpha_i$ is the normalized length of
each homopolymer chain.
Clearly, this Hamiltonian is again a convex function of the CS fields that is \emph{minimized} at the stationary SCFT solution.

Returning to the original model of a one-component homopolymer solution, the mean-field equations in full are
\begin{eqnarray}
  -i\left(-\partial_s -\nabla^2\right)\hat{\varphi}\left(\vect{r},s\right) & + & B\left(\Gamma\star\hat{\rho}\right)\left(\vect{r}\right)\hat{\varphi}\left(\vect{r},s\right)\nonumber\\&-&\sqrt{z}\delta\left(s-1\right)=0\\
  -i\left(\partial_s -\nabla^2\right)\varphi\left(\vect{r},s\right) & + & B\left(\Gamma\star\hat{\rho}\right)\left(\vect{r}\right)\varphi\left(\vect{r},s\right)\nonumber\\&+&i\sqrt{z}\delta\left(s-1\right)=0,
\end{eqnarray}
where $\Gamma(\vect{r})\star\hat{\rho}(\vect{r})$ is the convolution of the reduced Gaussian pair potential and the polymer-segment density operator. The only saddle-point solution for a system with periodic
boundary conditions is the following trivial, spatially homogeneous solution
which can be found analytically
\begin{eqnarray}
  \varphi\left(s\right) &=& \sqrt{z} \exp\left(-B \rho_M s\right)\Theta\left(s\right)\\
  \hat{\varphi}\left(s\right) &=& i\sqrt{z}\exp\left(-B \rho_M \left(1-s\right)\right)\Theta\left(1-s\right),
\end{eqnarray}
where $\Theta$ is the Heaviside function, and the mean-field reduced monomer density is
given as the solution of $\rho_M = z\exp\left(-B \rho_M\right)$.
This mean-field solution is of course identical to the SCFT analysis of the familiar
auxiliary-field representation of the model, for which one can find a purely imaginary
saddle-point solution for $w$:
\begin{eqnarray}
  Nw^\star & = & -i B \rho_M,\\
  \rho_M & = & z e^{-iNw^\star}.
\end{eqnarray}
%
\subsection{Complex Langevin Dynamics for the CS Formalism}
\label{sec:cl}
Developing viable numerical schemes for unapproximated simulation of polymer field theories
requires care in handling the \emph{sign problem} that arises from complex actions.
In this regard, the CS formalism is no different: straightforward integration
over real-valued CS field configurations, e.g., by Monte Carlo sampling methods,
is prohibitively inefficient due to integrands that oscillate strongly in sign.
One strategy for ameliorating this sign problem would be to deform the
integration path into the complex plane to pass through a saddle point on a
near-constant-phase trajectory.
The action functional is everywhere analytic, so deformations of this type pose
no immediate problems.
However, we have found that the complex Langevin
(CL) dynamics scheme\cite{Parisi,Ganesan2001,Glenn2006} is a more useful technique for simulating
complex-valued (AF) polymer field theories, since it is adaptive and does not require saddle points
or constant phase contours to be computed in advance.
However, as described below, additional
complications arise in devising a thermodynamically consistent CL scheme for the CS framework.

A first attempt to develop a complex Langevin sampling scheme for the CS model might
involve a ``diagonal descent'' ficticious dynamics.
In such a scheme, the fields are again promoted to complex numbers, and
each field mode is relaxed according to the instantaneous steepest-descent force while
simultaneously driven by a Langevin noise that is chosen to produce the correct importance-sampled distribution
of field configurations.
Such a scheme would have equations of motion of the form
\begin{eqnarray}\label{f21}
  \partial_t \varphi &=& - \frac{\delta H\left[\varphi,\hat{\varphi}\right]}{\delta \varphi\argrs} + \mu\left(\vect{r},s,t\right)\\
  \label{eqn:eomdd1}
                     &=&-i\left[\partial_s +\nabla^2-B\Gamma(\vect{r})\star\hat{\rho}(\vect{r})\right]\hat{\varphi}\argrs\nonumber\\
                     &+&\sqrt{z}\delta\left(s-1\right)+ \mu\left(\vect{r},s,t\right)\\
  \partial_t \hat{\varphi} &=& - \frac{\delta H\left[\varphi,\hat{\varphi}\right]}{\delta \hat{\varphi}\argrs} + \hat{\mu}\left(\vect{r},s,t\right)\\
  \label{eqn:eomdd2}
                         &=&i\left[\partial_s -\nabla^2+B\Gamma(\vect{r})\star\hat{\rho}(\vect{r})\right]\varphi\argrs\nonumber\\
                         &-&i\sqrt{z}\delta\left(s\right) + \hat{\mu}\left(\vect{r},s,t\right),
\end{eqnarray}
where in this case $\mu$ and $\hat{\mu}$ are real-valued Gaussian-distributed random numbers.

By eliminating the Langevin noise from Eqn. \ref{eqn:eomdd1} and \ref{eqn:eomdd2} and seeding at $t=0$ with randomly generated
fields that are purely real for $\varphi$ and purely imaginary for $\hat{\varphi}$, these equations of motion do
indeed relax to recover a physical saddle-point solution at long time.
Unfortunately, the diagonal descent scheme --- even in the absence of Langevin noise ---
is numerically ill-behaved, exhibiting marginal stability and
oscillatory trajectories to the stationary solution, as shown in Fig.~\ref{fig:SCFT}(a-c).
While the ultimate stationary solution is consistent with the exact mean-field value,
the strongly oscillating trajectory would
certainly be unsuitable for obtaining more complicated inhomogeneous SCFT solutions
or as a basis for stochastic CL simulations.

A dramatic improvement in stability is obtained by applying a specific non-Hermitian mobility
matrix to the field relaxation equations:
\begin{eqnarray}\label{eqn:dynamicalscheme}
\partial_t\left(\begin{array}{c}
\varphi\\\hat{\varphi}
\end{array}
\right)=-\left(\begin{array}{cc}
0&i\\i&0
\end{array}
\right)
\left(\begin{array}{c}
\delta H/ \delta\varphi \\ \delta H/ \delta\hat{\varphi}
\end{array}
\right)
+
\left(\begin{array}{c}
\mu \\ \hat{\mu}
\end{array}
\right).
\end{eqnarray}
This matrix has the effect of
decoupling the dynamics for $\varphi$ and $\hat{\varphi}$ to linear order, while also implicitly
including a Wick rotation so that the forcing term drives the dynamics to the mean-field configuration.
It is demonstrated in Fig.~\ref{fig:SCFT}(d), again without
noise applied, that Eqn.~\ref{eqn:dynamicalscheme} is a robust scheme for relaxing to SCFT solutions of the CS field theory.

\begin{figure*}[h!t]
\begin{center}
  {\includegraphics[bb=0 0 350 239, scale=1.2,draft=false]{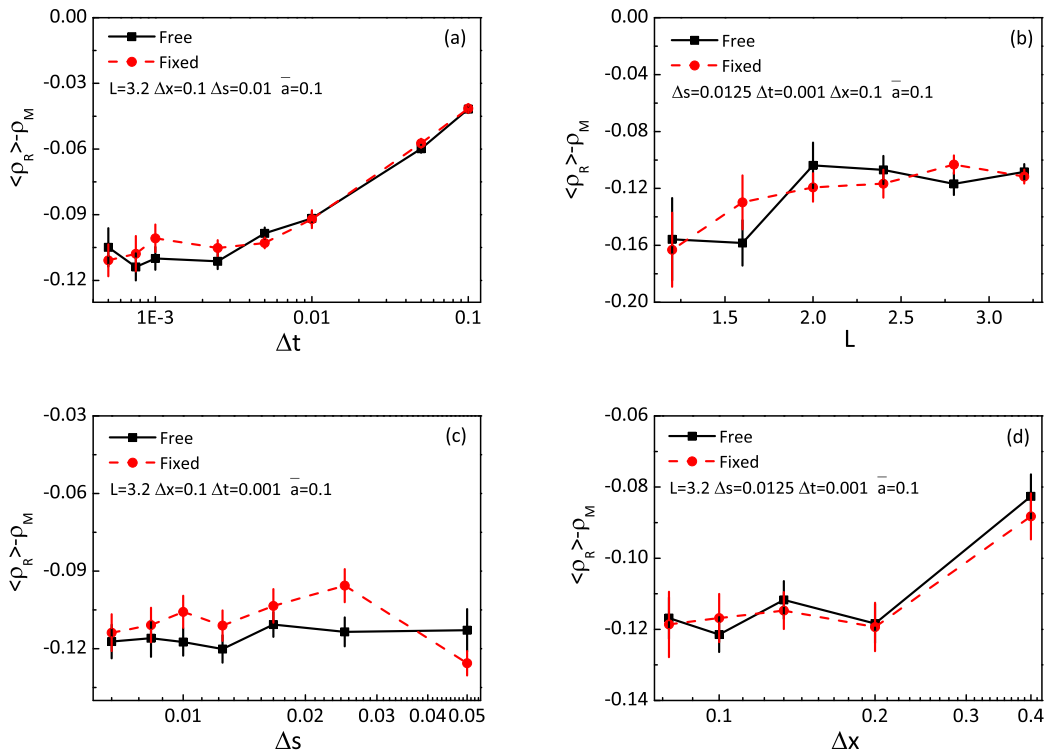}}
\caption{%
The dependence of the real part of the spatially and thermally
averaged polymer density, $\left<\rho_R\right>$, with the corresponding saddle-point values $\rho_M$
removed, on (a) the iteration time step $\Delta t$, (b) simulation cell size $L$, (c) contour discretization $\Delta s$
and (d) space discretization $\Delta x$ for both \emph{free boundary} and \emph{fixed boundary} cases.
All calculations were conducted for $B=1$ and $\sqrt{z}=1$.
The error bars displayed are one standard deviation of the error of the mean.
}
\label{fig:numera}
\end{center}
\end{figure*}
Beyond mean-field theory, the Gaussian noise terms $\mu$ and $\hat{\mu}$ must be carefully constructed in order to
produce a CL dynamics scheme that is \emph{thermodynamically consistent};
namely that has a steady-state probability distribution of field configurations consistent with the complex
Boltzmann weight of the partition function in Eqn.~\ref{eqn:GCECS}.
We provide a proof in Appendix \ref{sec:csnoise} that thermodynamic consistency in the present case imposes
restrictions on the second moments of the real (R) and imaginary (I) components of the noise terms
(all first moments should vanish):
\begin{eqnarray}\label{f23}
\left<\mu^2_{R}\right>-\left<\mu^2_{I}\right> & = & 0\\
\left<\hat{\mu}^2_{R}\right>-\left<\hat{\mu}^2_{I}\right> & = & 0 \\
\left<\mu_{R}\mu_{I}\right> & = & 0\\
\left<\hat{\mu}_{R}\hat{\mu}_{I}\right> & = &0\\
  \label{eqn:noiseconstraint5}
\left<\mu\left(\vect{r},s,t\right)\hat{\mu}\left(\vect{r}^\prime,s^\prime,t^\prime\right)\right>
&=&i \delta\left(\vect{r}-\vect{r}^\prime\right)\delta\left(s - s^\prime\right)\delta\left(t-t^\prime\right).\nonumber\\
\end{eqnarray}
These constraints on the noise correlations are required for Eqn.~\ref{eqn:dynamicalscheme} to generate the appropriate ensemble of
field configurations.
In particular, notice that Eqn.\ \ref{eqn:noiseconstraint5} demands that the noise applied be both complex valued
and correlated between the $\varphi$ and $\hat{\varphi}$ equations of motion --- a
consequence of the complex mobility matrix that was introduced for stabilizing the dynamics.
Within the auxiliary-field framework of this model, it is possible to attain thermodynamic
consistency using only real-valued noise applied to the single equation of motion for $w$\cite{Glenn2006}.

The restrictions placed on the second moments of the random variables
are insufficient to uniquely specify the noise statistics.
While all choices consistent with the above constraints should yield identical time-averaged properties,
we expect that non-observable properties that strongly affect sampling efficiency,
such as population variance and correlation times, can vary appreciably between different choices.
Among a variety of schemes investigated, our best-performing CL scheme (Appendix \ref{sec:csnoise})
employs two \emph{real and independent} Gaussian
random deviates, $\eta_1$, $\eta_2$, with vanishing first moments, and second moments given by
$\left<\eta_i\left(\vect{r},s;t\right) \eta_j\left(\vect{r}^\prime, s^\prime, t^\prime\right)\right>
= 2\delta\left(\vect{r}-\vect{r}^\prime\right)\delta\left(s-s^\prime\right)\delta\left(t-t^\prime\right)\delta_{i,j}$,
applied to Eqn.~\ref{eqn:dynamicalscheme} using the transformation
$\mu_R = \eta_1/\sqrt{2}$, $\mu_I = \eta_2/\sqrt{2}$, $\hat{\mu}_R = \eta_2/\sqrt{2}$, and $\hat{\mu}_I = \eta_1/\sqrt{2}$.
Our CS-CL dynamics for this model can be summarized as:
\begin{eqnarray}\label{f25}
  \partial_t\varphi\left(\vect{r},s,t\right)&=&-i\frac{\delta H\left[\varphi,\hat{\varphi}\right]}{\delta\hat{\varphi}\left(\vect{r},s,t\right)}+\frac{\left(\eta_1\left(\vect{r},s,t\right)+i\eta_2\left(\vect{r},s,t\right)\right)}{\sqrt{2}}\nonumber\\
  \partial_t\hat{\varphi}\left(\vect{r},s,t\right)&=&-i\frac{\delta H\left[\varphi,\hat{\varphi}\right]}{\delta\varphi\left(\vect{r},s,t\right)}+\frac{\left(\eta_2\left(\vect{r},s,t\right)+i\eta_1\left(\vect{r},s,t\right)\right)}{\sqrt{2}}.\nonumber\\
\end{eqnarray}

Finally, we have explored two approaches for tackling the chain-end delta-function sources present in the
force terms of the equations of motion.
In the first method, we permit the boundary components $\varphi\left(\vect{r},s=0\right)$ and $\hat{\varphi}\left(\vect{r},s=1\right)$ to fluctuate, subject
to Langevin noise.
Without the noise, these modes will dynamically relax to $\sqrt{z}$ and $i\sqrt{z}$ respectively.
We denote the method with fluctuating chain-end initial conditions the ``free boundary'' approach.
Care must be taken to treat the discretized form of the delta-function source and the $\partial_s$ operator to ensure causal response fields.
This aspect is discussed in Appendix \ref{sec:discrete}.
The second method, which we label ``fixed boundary'' has a time-independent value for
$\phi\left(\vect{r},s=0\right)$ and $\hat{\varphi}\left(\vect{r},s=1\right)$.
The specific value of the fields adopted at the ``fixed boundary'' are determined by removing the noise
and allowing the fields to relax during the thermalization
to a static value that approaches (in the limit $\Delta s \rightarrow 0$) $\sqrt{z}$ for $\varphi\left(s=0\right)$
and $i\sqrt{z}$ for $\hat{\varphi}\left(s=1\right)$.
This scheme ensures consistency between the discretized form of the source terms and $\partial_s$.
In contrast, fixing the field boundary values to their continuum limit results in a $\Delta s$ dependent bias that
is removed as $\Delta s \rightarrow 0$.
It can be demonstrated by integration over a small (vanishing) contour slice and deriving a thermodynamically consistent stochastic boundary condition on the chain-end
field contributions that the two methods should be equivalent.
We test this numerically in the following section.


\section{CL Results}
\label{sec:results}
\subsection{Numerical Convergence of Fluctuating Observables}
\label{sec:convergence}
\begin{figure*}[h!t]
\begin{center}
  {\includegraphics[bb=0 0 350 239, scale=1.2,draft=false]{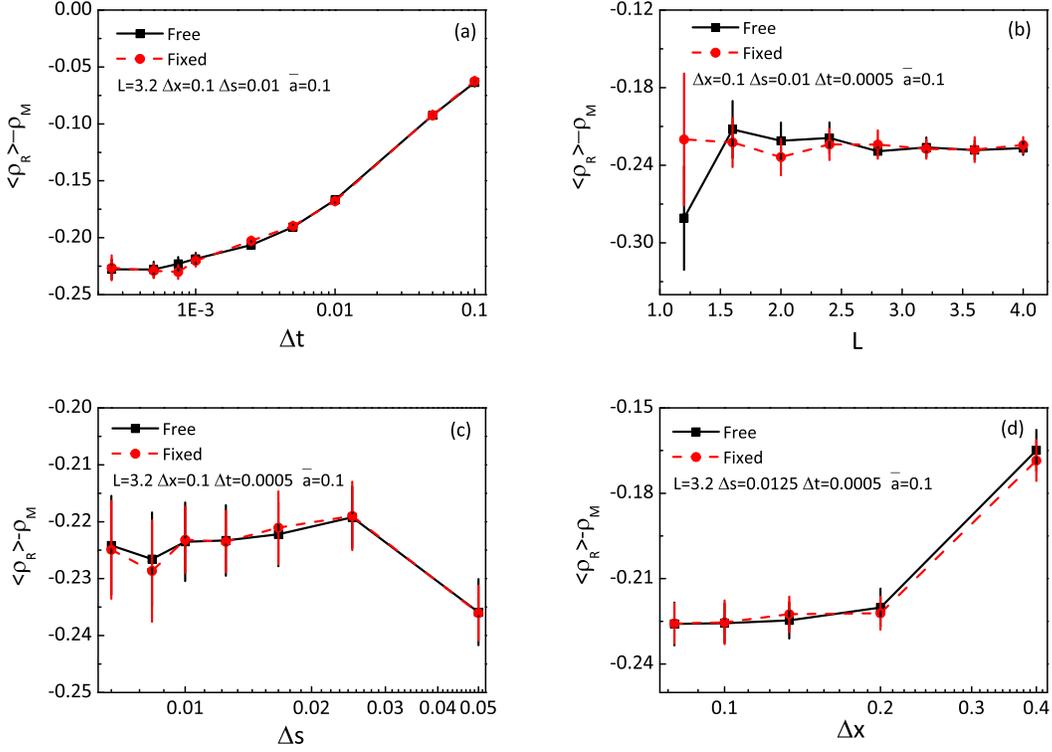}}
\caption{As Fig.~\ref{fig:numera} but for $B=2$ and $\sqrt{z}=10$.
The dependence of $\left<\rho_R\right>-\rho_M$ on
$L$, $\Delta s$ and $\Delta x$ are largely the same as in Fig.~\ref{fig:numera}.}
\label{fig:numerb}
\end{center}
\end{figure*}

We first investigate the systematic elimination of
discretization and finite-size errors in our simulations.
We employ a cubic simulation cell with periodic boundary conditions and
a pseudospectral semi-implicit solver detailed in Appendix \ref{sec:discrete}.
Our observable of choice is the real part of the spatially averaged polymer chain
density, $\left<\rho_R\right>$, with the mean-field contribution, $\rho_M$, removed.
Convergence studies are shown for a lower density, weakly interacting solution in Fig.~\ref{fig:numera},
and for a higher density, more strongly interacting solution in Fig.~\ref{fig:numerb}.
For both cases, we choose a model with interaction range parameter $\bar{a}=0.1\,R_g$.
We note that larger $B$ and $z$ parameters are accessible with our algorithms, but
the small time step required makes calculations in this limit more demanding.

Our studies show that a time step sufficient to achieve stable and accurate field trajectories for
this model using our semi-implicit time-integration algorithm is $\Delta t\leq 0.0005$.
Figs.~\ref{fig:numera}(b) and \ref{fig:numerb}(b) show that the simulation cell size $L$ must be
large enough to eliminate finite-size bias from truncating below the solution correlation length.
We find that $L\geq2.8\,R_g$ is always sufficient in the present case.
Figs.\ \ref{fig:numera}(c,d) and \ref{fig:numerb}(c,d) suggest that $\Delta s\leq 0.02$ and $\Delta x\leq 0.2$
are required to accurately resolve contour and spatial sampling.
Note that the $\Delta x \rightarrow 0$ limit is well defined, due to the regularization of interactions in the model.
It should be noted that both Fig.~\ref{fig:numera} and Fig.~\ref{fig:numerb} demonstrate statistical consistency
between CS-CL conducted with ``free boundary'' and ``fixed boundary'' conditions on the contour dependence of the
CS fields.
Hence, as anticipated, the fluctuations at the contour boundary are here demonstrated to be negligable, and
one may simulate with a fixed contour boundary condition determined by the source term of the action.
\subsection{Comparison of CS and AF observables}
\label{sec:rhoresults}
We now demonstrate that our numerical implementation of the CS-CL sampling scheme
has steady-state time averages that correspond to the correct ensemble averages.
All calculations were conducted in a cell of side
length $L=3.2\,R_g$, which eliminates finite-size errors, and with
$\Delta x=\bar{a}=0.1\,R_g$, $\Delta s=0.01$, and $\Delta t=0.0005$.

\begin{figure}[h]
\begin{center}
  {\includegraphics[bb=0 0 360 250, scale=0.62,draft=false]{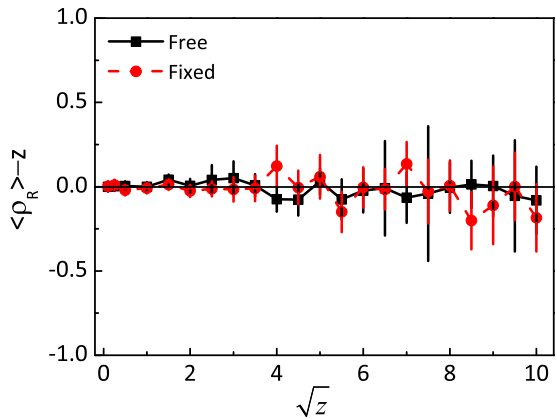}}
\caption{For $B=0$, the ideal solution, comparison of the values
of the real part of the spatially and temporally averaged fluctuation-enhanced polymer density, $\left<\rho_R\right>-z$,
where $\rho_M=z$ is the exact value in both mean-field and fully fluctuating limits.
Statistical consistency between the results from CS-CL simulations and the analytical study is shown in the figure.
For both \emph{free boundary} and \emph{fixed boundary} method, calculations were
conducted for $\Delta x=\bar{a}=0.1 \: R_g$, $L=3.2 \: R_g$, $\Delta s=0.01$, and $\Delta t=0.0005$.
The error bars have the same meaning as in Fig.~\ref{fig:numera}}
\label{fig:ideal}
\end{center}
\end{figure}

We first analyse the ideal polymer solution ($B=0$), for which analytical studies can be conducted.
The spatially averaged polymer chain density $\left<\rho_R\right>$ for both fluctuating
and mean-field calculations should equal the chain activity, $z$.
The dependence of $\left<\rho_R\right>$ on $z$ is shown in Fig.~\ref{fig:ideal},
where the values of $\left<\rho_R\right>-z$ are always statistically equal to $0$
for $\sqrt{z}=0.1$--$10$, indicating agreement between numerical and exact results.
Furthermore, ``free boundary'' and ``fixed boundary'' simulations are again in agreement.

\begin{figure}[h]
\begin{center}
  {\includegraphics[bb=0 0 375 250, scale=0.62,draft=false]{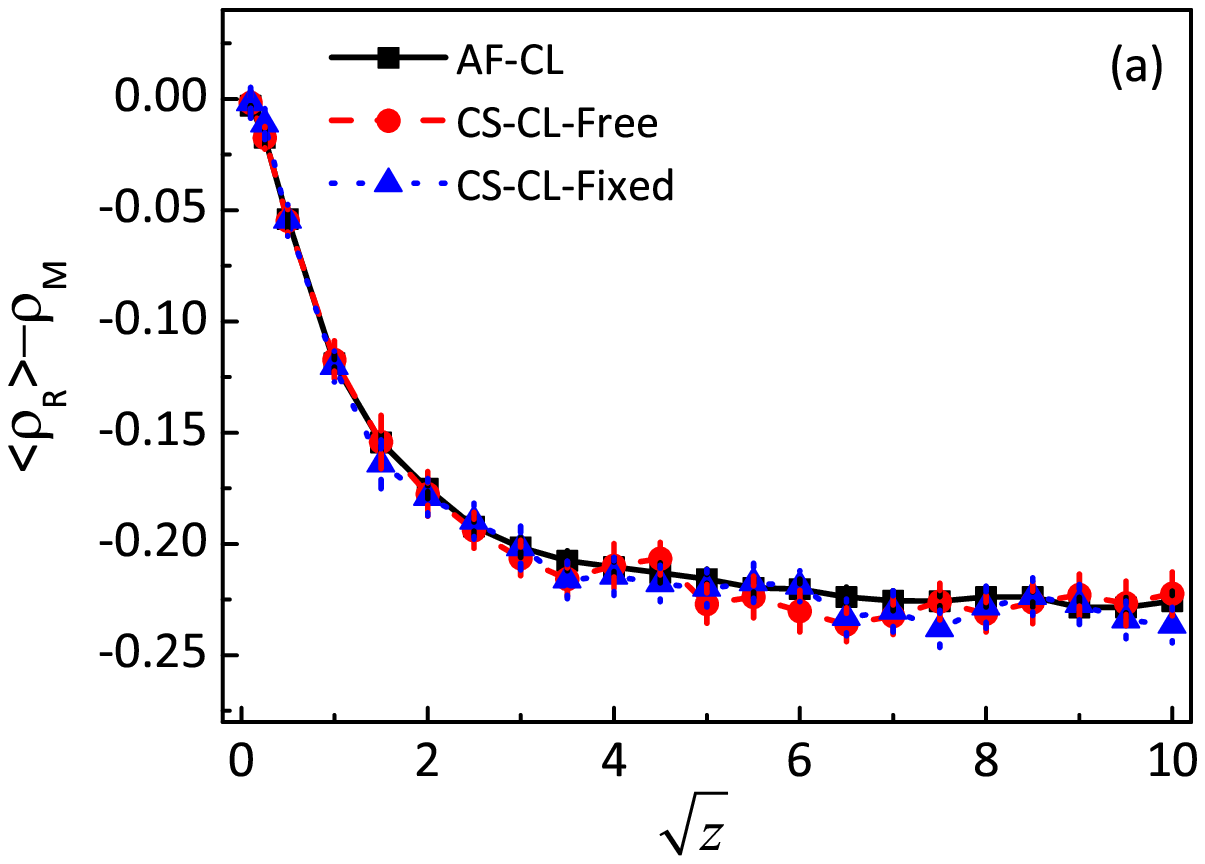}}
  {\includegraphics[bb=0 0 375 250, scale=0.62,draft=false]{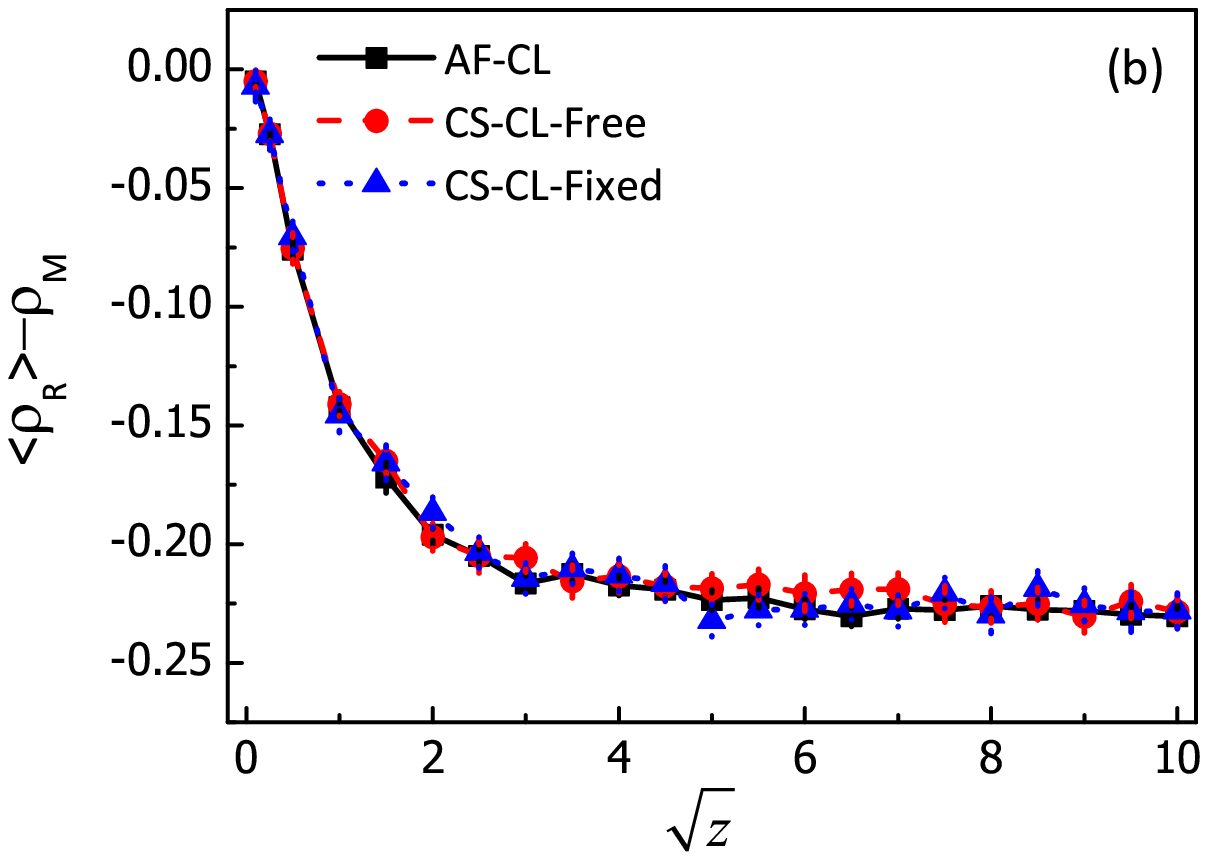}}
\caption{Comparison of the values of the real part of the spatially and temporally averaged polymer density, $\left<\rho_R\right>$,
which have been subtracted from their corresponding saddle-point values, $\rho_M$, for different values of the chain activity, $z$.
Statistical consistency is seen between the results from simulations based
on Edwards' auxiliary field representation and the new coherent states formalism.
Calculations were conducted for $B=1$ in (a); $B=2$ in (b).
All other parameters are $\Delta x=\bar{a}=0.1 \: R_g$, $L=3.2 \: R_g$, $\Delta s=0.01$, and $\Delta t=0.0005$.
The error bars have the same meaning as in Fig.~\ref{fig:numera}.}
\label{fig:CL}
\end{center}
\end{figure}

For $B\neq0$, exact analytic results are not available for the fully fluctuating theory.
For reference, we conducted simulations of this regularized polymer solution
model, with identical parameters, using the conventional AF-CL framework\cite{Villet}.
Fig.~\ref{fig:CL} shows the fluctuation contribution to the spatially averaged polymer
density, $\left<\rho_R\right>$, relative to the mean-field value $\rho_M$, for a range of
solution concentrations.
Two values of the excluded-volume interaction parameter have been simulated: $B=1$ (Fig.~\ref{fig:CL}(a))
and $B=2$ (Fig.~\ref{fig:CL}(b)).
The statistical consistency  between simulation results for the AF-CL and CS-CL methods demonstrates the
validity of our chosen complex Langevin dynamics.
\subsection{Causal Nature of the Green Function}
\label{sec:Gfn}
%
We now return to the analysis of the causal nature of our sampling scheme, which is required
for the functional integral denominator in Eqn.\ \ref{eqn:GtoCS} to be a field-independent constant.
The denominator is
$\int \mathcal{D}\varphi \int\mathcal{D}\hat{\varphi}
\exp\left(i\int d\vect{r}\int d\vect{r}^\prime \int ds \int ds^\prime\,
\hat{\varphi}\left(\vect{r}^\prime,s^\prime\right)\mathcal{L}\varphi\left(\vect{r},s\right)\right)$, which
is a standard Gaussian integral that can be related to the inverse of the determinant of $\mathcal{L}$.
Since $\det G$ is related to $\left(\det \mathcal{L}\right)^{-1}$, the requirement for the denominator to be a constant
independent of $iw\left(\vect{r}\right)$ can be recast as a requirement that $\det G$ is constant with respect to all
parameters.

The Green function that can be computed for a solution of interacting chains
is different from the Green function introduced in Section \ref{sec:theory} because
the latter is for non-interacting chains subject to an arbitrary imaginary external field, $iw\left(\vect{r}\right)$.
However, a Green function for the interacting solution can be thought of as a
one for a non-interacting solution integrated over a specific collection of
external field configurations, and so a $B$-independent constant for the
determinant of the interacting $G$ is consistent with a field-independent constant for
the determinant of the non-interacting $G$.

\begin{figure}[h]
\begin{center}
  {\includegraphics[bb=0 0 340 250, scale=0.62,draft=false]{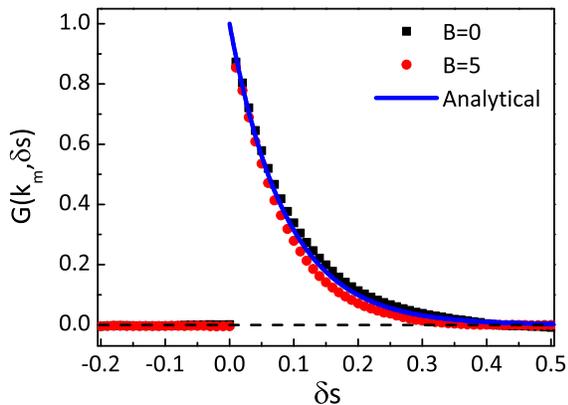}}
\caption{The dependence of the Green function on $\delta s$ for $k_m=\frac{2\pi}{L}$, $B=0$ and $B=5$ are
shown in the figure. The dashed line is used to guide the zero value.
All calculations were conducted for $\Delta x=0.2 \: R_g$, $L=3.2 \: R_g$, $\Delta s=0.01$, $\bar{a}=0.1 \: R_g$,
$\Delta t=0.0005$, and $\sqrt{z}=1$.}
\label{fig:GFn}
\end{center}
\end{figure}

First, we consider the ideal solution ($B=0$).
In this case, $G$ has an available closed-form expression:
\begin{eqnarray}
  G\left(\vect{r}-\vect{r}^{\prime},s,s^{\prime}\right)=\frac{1}{4\pi\left(s-s^{\prime}\right)^{\frac{3}{2}}}e^{-\frac{\left|\vect{r}-\vect{r}^{\prime}\right|^2}{4\left(s-s^{\prime}\right)}}\Theta\left(s-s^{\prime}\right),
\end{eqnarray}
with Fourier transform
\begin{eqnarray}\label{kgreen}
G\left(\vect{k},s,s^{\prime}\right)=e^{-k^2\left(s-s^{\prime}\right)}\cdot\Theta\left(s-s^{\prime}\right).
\end{eqnarray}
Numerically, the most convenient way to compute $G$ for a homogeneous (translationally invariant) system
is from the Fourier transform of Eqn.\ \ref{eqn:Gfn}:
\begin{eqnarray}
G\left(\vect{k},s,s^{\prime}\right)=-iV\left<\varphi\left(\vect{k},s\right)\hat{\varphi}\left(-\vect{k},s^{\prime}\right)\right>_0.
\end{eqnarray}
We note that a direct calculation of the $\vect{k}=0$ contribution to $G$ is spurious due to the
periodic boundary conditions employed in our simulations.
We thus study the Green function for the minimum non-zero $k_m={2\pi}/{L}$.
The dependence of $G\left(k_m,\delta s\right)$ on
$\delta s=s-s^{\prime}$ is shown in Fig.~\ref{fig:GFn}.
For the ideal solution, we find good agreement between the numerically computed $G$ and the analytic expression.
%
%
Both non-interacting and interacting solutions correctly obey the identities
$G\left(k,\delta s=0^{-}\right)=0$ and $G\left(k,\delta s=0^{+}\right)\rightarrow 1$.
Using our discretization approach (Appendix \ref{sec:discrete}), the Green function
has $G(k,0)=0$ for all $k$, for both interacting and non-interaction solutions.
A subtle feature of the discretization scheme is that $\varphi\left(s\right)$ is
correlated with $\hat{\varphi}\left(s-\Delta s\right)$ due to the first-order
$\partial_s$ operator acting in different directions on the two fields.
Therefore, the matrix representation of $G$ is related to that of $\mathcal{L}$
by an index shift in $s-s^\prime$, such that numerically $\det{\mathcal{L}}$ is a constant
that approaches $1$ in the limit $\Delta s\rightarrow 0$.
This statement applies for all non-zero $\vect{k}$.
This numerical demonstration gives \emph{a posteriori} confirmation
of the validity of neglecting the denominator in Eqn.\ \ref{eqn:GtoCS}.
%
\subsection{Comparitive Efficiency of CS-CL and AF-CL Numerical Schemes}
\label{sec:efficiency}
We now discuss the relative efficiency of AF-CL and CS-CL methods.
The former is implemented using a standard pseudospectral solver with
exponential time differencing to advance the CL equations\cite{VilletThesis}.
The latter is implemented with the numerical scheme presented in the present paper.

From the perspective of computational efficiency, the crucial metric involves quantifying the
accuracy with which one can make predictions of thermal properties within a
given simulation time, as measured by the magnitude of the standard error of the
estimate of the mean of that observable (i.e., the error bars plotted in Figs.~\ref{fig:numera}--\ref{fig:CL}).
The standard error decreases as $1/\sqrt{N}$ as the number of CL time steps, $N$, is increased,
corresponding to an improved confidence that the sample mean is equal to the
population mean (i.e., the correct value of the observable), assuming the central limit theorem holds.
The standard error of the mean is related linearly to the sample standard deviation, and both
quantities should be corrected for biases resulting from finite sample size
and Markovian serial correlation.

\begin{figure}[h]
\begin{center}
  {\includegraphics[bb=0 0 340 250, scale=0.62,draft=false]{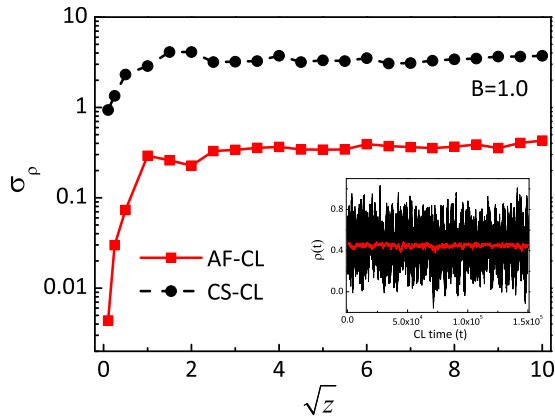}}
\caption{Serial-correlation-corrected sample standard deviation of the monomeric density operator calculated
  over long ($150000$ time step) CS-CL and AF-CL simulations with $B=1.0$ and various $z$, corresponding to simulations
reported in Fig.~\ref{fig:CL}(a). The population
standard deviation of the distributions sampled differ by approximately an order of magnitude
with CS-CL the larger. Inset: example time-dependent trace of the monomeric density operator
for $B=1.0$, $z=1.0$, demonstrating the larger fluctuations in the CS-CL case.}
\label{fig:sd}
\end{center}
\end{figure}

Our simulations show that the CS-CL scheme developed and presented in this manuscript stochastically
samples distributions of observables with a vastly increased population standard deviation compared to
the AF-CL method (see Fig.~\ref{fig:sd}).
This observation is quite intuitive: the discretized CS-CL dynamics has $2MN_s$ modes driven by noise (where
$M$ is the number of plane waves and $N_s=\frac{1}{\Delta s}$ is the number of samples in the contour variable), and
the noise amplitude scales as $1/\sqrt{\Delta s}$.
In contrast, the AF-CL dynamics applies noise only to $M$ degrees of freedom.
On the other hand, due to the increased noise application, the CS-CL dynamical operators exhibit shorter
correlation times (typically by $\sim2\times$ in the simulations reported here).
The compound effect on the correlation-corrected sample standard deviations is reported in Fig.~\ref{fig:sd} for
a set of grand-canonical solutions with excluded-volume interaction $B=1.0$.

\begin{figure}[h]
\begin{center}
  {\includegraphics[bb=0 0 342 250, scale=0.62,draft=false]{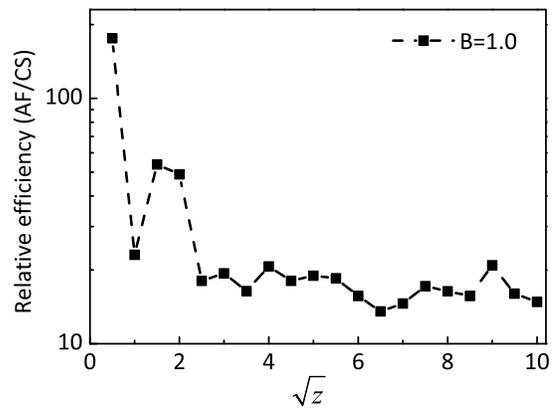}}
\caption{Relative efficiency of AF-CL simulations over CS-CL simulations, as measured by the factor reduction in number of simulation steps required to produce errors of the mean of a specified magnitude. This quantity is equal to the squared ratio of the
error of the mean of $\bar{\rho}$ when both methods are used with an equal number of CL time steps ($150,000$).}
\label{fig:efficiency}
\end{center}
\end{figure}

Finally, the relative magnitude of the standard deviations can be translated into a relative efficiency of the
two methods, as measured by the \emph{square} of the ratio of the standard deviations of the
errors of the mean (effectively the squared ratios of the AF-CL and CS-CL error bars reported in Fig.~\ref{fig:CL}) for
a fixed number of CL time steps.
This quantity, shown in Fig.~\ref{fig:efficiency} determines the relative number of CL simulation steps
required for the CS method to produce error bars of the same magnitude as the AF method.
In the present case, we find simulations $\sim 20\times$ longer would be required for CS-CL to match the accuracy of AF-CL.
This explains why the error bars in Fig.~\ref{fig:CL} are $\sim4$--$5\times$ larger for CS-CL than for AF-CL, with
all simulations completing the same number of time steps.

\section{Future Prospects}
We have introduced a stable and efficient numerical scheme for sampling the fully fluctuating CS field theory of
the grand-canonical ensemble of solvated homopolymers.
The efficiency of the method still lags the more mature AF-CL approach, as demonstrated in
Section \ref{sec:efficiency}.
It is therefore crucial that future algorithmic developments address modifications to the sampling scheme
such that either the population variance of sampled operators or their correlation time is reduced.
For the latter, one option would involve development of more advanced time-stepping algorithms
with improved stability \emph{and accuracy} that would permit acceleration of the dynamical trajectories.
Based on convergence data shown in Fig.~\ref{fig:numera}(c) and \ref{fig:numerb}(c), we suggest that higher-order
methods for sampling the contour variable would be of limited value for \emph{fluctuating} simulations, at least for
polymer models with similar sampling requirements.
With further algorithmic developments, we anticipate that the many potential advantages of the
coherent-states framework will be realized in practice.

We now discuss additional physical aspects of the theory that are not augmented by simulation data, and could form the
basis of future studies.
\subsection{Transformation to the Canonical Ensemble}
\label{sec:CE}
For many applications of interest, it is more convenient to work with the
canonical ensemble rather than the grand canonical ensemble.
For example, for forced stabilization of an interface in an immiscible blend of components,
or for simulations in which precise control of component concentrations is desired.
In this section we illustrate the route for transforming to such an ensemble while
remaining within the CS framework.
We begin with the standard relationship between the partition functions of grand
canonical ($\Xi\left(z,V,T\right)$) and canonical ($Z_C\left(n,V,T\right)$) ensembles with
a focus on the homopolymer solution model used throughout this paper:
\begin{eqnarray}\label{f12}
\Xi\left(z,V,T\right)=\sum^\infty_{n=0} z^n Z_C\left(n,V,T\right).
\end{eqnarray}
The inverse transformation is
\begin{eqnarray}\label{f13}
  Z_C\left(n,V,T\right)=\frac{1}{n!}\left.\frac{\partial^n \Xi\left(z,V,T\right)}{\partial z^n}\right|_{z=0}.
\end{eqnarray}
Inserting the CS form of the grand partition function (Eqns.\ \ref{eqn:GCECS} and \ref{eqn:CSaction_dimless}), we find
\begin{eqnarray}\label{f14}
Z_C\left(n,V,T\right)=\frac{V^n}{\left(n!\right)^2}\int \mathcal{D}\hat{\varphi}\int \mathcal{D}\varphi\, \exp\left(-H\left[\hat{\varphi},\varphi\right]\right),
\end{eqnarray}
where the canonical Hamiltonian is
\begin{eqnarray}\label{f15}
H\left[\hat{\varphi},\varphi\right]&=&-i\int^{1}_{0} ds\int d\vect{r} \, \hat{\varphi}\left(\vect{r},s\right)
  \left(\partial_s-\nabla^2\right)\varphi\left(\vect{r},s\right)\nonumber\\
  & +&\frac{B}{2}\int d\vect{r}\int d\vect{r}^{\prime}\,\hat{\rho}(\vect{r})
  \Gamma \left(|\vect{r}-\vect{r}^{\prime} | \right)
  \hat{\rho}\left(\vect{r}^{\prime}\right)\nonumber\\
  & -&n \ln\left(\frac{-i}{\sqrt{V}}\int d\vect{r}\,\hat{\varphi}\left(\vect{r},0\right)\right)\nonumber\\
  &-&n \ln\left(\frac{1}{\sqrt{V}}\int d\vect{r}\,\varphi\left(\vect{r},1\right)\right),
\end{eqnarray}
where the polymer chain density operator $\hat{\rho}\left(\vect{r}\right)$ is as defined previously in Eqn.\ \ref{f10}.
It is interesting to note that only the source terms of the Hamiltonian differ between canonical and grand-canonical ensembles; in
each case the appropriate assembly of polymer chains is generated by the specific form of the source terms.
As in the auxiliary-field framework, the action functional contains natural logarithms, which modify the analytic
structure of the functional by introducing a branch cut.
This feature usually has no detrimental consequences in numerical simulations.

\subsection{Assemblies of Asymptotically Long Polymer Chains}
\label{sec:GSD}
One of the utilities of the CS formalism is to investigate fluctuation phenomena in
assemblies of asymptotically long interacting polymer chains --- simulations
that would be extremely computationally demanding in the AF framework.
One can conduct such studies by making the ground-state-dominance approximation (GSD).
We note that GSD approximation is only valid when polymers of high molecular weight are bound
to regions with dimensions comparable to $R_g$, so that the spectrum of eigenstates of
$\mathcal{L}$ has a finite gap between the ground state and the first excited state.
An example might be the adsorption of long polymer chains to a substrate.

We proceed by expanding the CS fields in the complete basis of eigenstates of $\mathcal{L}$
\begin{eqnarray}\label{f16}
\varphi\argrs&=&\frac{1}{\sqrt{N}}\sum_{K}e^{-\mu_K s}\psi_K(\vect{r})\nonumber\\
\hat{\varphi}\argrs&=&\frac{1}{\sqrt{N}}\sum_{K}e^{\mu_K s}\hat{\psi}_K(\vect{r})
\end{eqnarray}
where $\mu_K$ is $K^\mathrm{th}$ eigenvalue, while $\psi_K(\vect{r})$ and $\hat{\psi}_K(\vect{r})$ are the
corresponding eigenstates.
In the GSD approximation, which is valid in the limit of very
large $N$ and a non-degenerate spectrum, we retain only the lowest eigenstate, $\mu_m$:
\begin{eqnarray}
  \label{eqn:GSD}
  \varphi\argrs&\approx&\frac{1}{\sqrt{N}}e^{-\mu_m s}\psi(\vect{r}), \\
  \label{eqn:GSDhat}
  \hat{\varphi}\argrs&\approx&\frac{1}{\sqrt{N}}e^{\mu_m s}\hat{\psi}(\vect{r}).
\end{eqnarray}
Inserting Eqns.\ \ref{eqn:GSD} and \ref{eqn:GSDhat} into Eqn.\ \ref{eqn:CSaction_dimful} yields
\begin{eqnarray}\label{eqn:HGSDint}
  H[ \hat{\psi},\psi ] &=&-i\int d\vect{r} \,\hat{\psi}\left(\vect{r}\right)\left(-\mu_m-\frac{b^2}{6}\nabla^2\right)\psi\left(\vect{r}\right)\nonumber\\
& +&\frac{1}{2}\int d\vect{r}\int d\vect{r}^{\prime}\,\hat{\rho}\left(\vect{r}\right)u\left(\left|\vect{r}-\vect{r}^{\prime}\right|\right))\hat{\rho}\left(\vect{r}^{\prime}\right)\nonumber\\
& -&\sqrt{\frac{z^\prime }{N}}\int d\vect{r}\,\left[-i\hat{\psi}(\vect{r})+e^{-\mu_m N}\psi(\vect{r})\right],
\end{eqnarray}
where the monomer density is now
\begin{eqnarray}\label{f19}
\hat{\rho}(\vect{r})=-i\hat{\psi}(\vect{r})\psi(\vect{r}).
\end{eqnarray}
Taking $N\rightarrow\infty$, the source term in Eqn.~\ref{eqn:HGSDint} disappears
resulting in a CS-GSD action functional, which is essentially a Ginzburg-Landau theory of
long polymer chains\cite{GarelOrlandPitard1997}
\begin{eqnarray}\label{f20}
  H[\hat{\psi},\psi ]&=&-i\int d\vect{r}\,\hat{\psi}\left(\vect{r}\right)\left(-\mu_m-\frac{1}{6}\nabla^2\right)\psi\left(\vect{r}\right)\nonumber\\
& +&\frac{B^\prime}{2}\int d\vect{r}\int d\vect{r}^{\prime}\,\hat{\rho}\left(\vect{r}\right) \Gamma \left(\left|\vect{r}-\vect{r}^{\prime}\right|\right)\hat{\rho}\left(\vect{r}^{\prime}\right),
\end{eqnarray}
where all lengths have been rescaled by the statistical segment length $b$, and $B^{\prime}=u_0/b^{3}$ is a
dimensionless excluded-volume interaction parameter.
Note that $\mu_m$ takes the role of a \emph{monomer} chemical potential, which is the only relevant quantity in
a system of asymptotically long chains, and the statistical segments are no-longer labeled by their position along
the polymer chain backbone.

\section{Conclusions}
\label{sec:Concl}
In this paper, we have introduced the first numerical study of a coherent states polymer field theory.
For this purpose, a complex Langevin scheme was devised that circumvents the sign problem, and is both stable and efficient.
A conventional diagonal-descent scheme is numerically ill-behaved.
However, we found that application of a non-Hermitian mobility matrix to the field relaxation equations,
coupled with appropriate complex noise statistics, yields CL simulations with stable trajectories and correct time-averaged properties.
Statistical consistency between CS-CL simulations and analytical solutions has been obtained for ideal polymer solutions, and between
CS-CL and AF-CL for non-ideal solutions.
The techniques presented in this paper should be applicable to deriving
theoretical and numerical schemes required to simulated other
polymeric systems in the CS framework.

In addition to the CS formalism for a grand canonical model, we presented the route for deriving
CS canonical ensemble theories, and a method for studying asymptotically long polymer chains with composition
fluctuations fully included using a simplified field theory in the ground-state-dominance approximation.

Future developments will require the identification of efficient numerical techniques
for reducing population variance of operators arising from field fluctuations in the CS
scheme to render it competitive with AF simulations.
We also plan to explore the use of the CS framework for conducting systematic coarse-graining and
numerical renormalization-group studies\cite{Villet2010}, for which the
locality and finite order of the action functional is expected to simplify the identification of suitable basis
functionals for building trial coarse-grained actions.

\section*{Acknowledgements}

XKM and KTD were supported by the National Science Foundation, respectively under
award DMR-CMMT-1160895 and the SOLAR program award No.\ CHE-1035292.
This work was partially supported by the MRSEC program of the National Science Foundation
under award No.\ DMR-1121053.  Computational resources were provided by the Center for
Scientific Computing at UCSB, a partnership between
CNSI and MRL with facilities provided by NSF award CNS-0960316.


\appendix
\section{Thermodynamic consistency in complex Langevin sampling of the CS equations}
\label{sec:csnoise}
In this appendix we demonstrate derivation of the appropriate noise correlations for a complex Langevin dynamics of
the form Eqn.\ \ref{eqn:dynamicalscheme}, with the stabilizing mobility matrix included, to reproduce the correct thermal averages.
This derivation follows Appendix D of Ref.\ \onlinecite{Glenn2006}.
The ensemble average of an operator $G[\varphi,\hat{\varphi}]$ in the CS field theory can be defined as follows
\begin{eqnarray}\label{a1}
\left\langle G\left[\varphi,\hat{\varphi}\right]\right\rangle=\int \mathcal{D}\varphi\int \mathcal{D}\hat{\varphi}~G\left[\varphi,\hat{\varphi}\right]P_C\left[\varphi,\hat{\varphi}\right]
\end{eqnarray}
where both $\varphi(\textbf{r},s)$ and $\hat{\varphi}(\textbf{r},s)$ are real-valued coherent
states (CS) and $P_C\left[\varphi,\hat{\varphi}\right]$ is a complex-valued Boltzmann weight defined by
\begin{eqnarray}\label{a2}
P_C\left[\varphi,\hat{\varphi}\right]=\frac{\exp\left(-H\left[\varphi,\hat{\varphi}\right]\right)}{\int \mathcal{D}\varphi\int \mathcal{D}\hat{\varphi}\,\exp\left(-H\left[\varphi,\hat{\varphi}\right)\right]}.
\end{eqnarray}
Though $P_C$ appears to play the role of a probability density, it is not positive-semi-definite for
$H\left[\varphi,\hat{\varphi}\right]$ complex.
We assume that there exists a real, non-negative probability density $P\left[\varphi,\hat{\varphi}\right]$,
where both $\varphi$ and $\hat{\varphi}$ are now \emph{complex-valued} fields with $\varphi_R$, $\varphi_I$
and $\hat{\varphi}_R$, $\hat{\varphi}_I$ corresponding to their real and imaginary parts separately, so
that the thermal averages in Eqn.\ \ref{a1} can be re-expressed as
\begin{eqnarray}\label{a3}
\left\langle G\left[\varphi_R,\hat{\varphi}_R\right]\right\rangle=\int \mathcal{D}\varphi_R\int \mathcal{D}\varphi_I\int \mathcal{D}\hat{\varphi}_R\int \mathcal{D}\hat{\varphi}_I \nonumber\\ G\left[\varphi_R+i\varphi_I,\hat{\varphi}_R+i\hat{\varphi}_I\right]P\left[\varphi_R,\varphi_I,\hat{\varphi}_R,\hat{\varphi}_I\right].
\end{eqnarray}
Direct comparison of the right-hand sides of Eqns.~\ref{a1} and \ref{a3} indicates that they are equivalent
if a $P\left[\varphi_R,\varphi_I,\hat{\varphi}_R,\hat{\varphi}_I\right]$ exists such that
\begin{eqnarray}\label{a4}
& &P_C\left[\varphi_R,\hat{\varphi}_R\right]=\nonumber\\
& &\int \mathcal{D}\varphi_I\int \mathcal{D}\hat{\varphi}_I ~P\left[\varphi_R-i\varphi_I,\varphi_I,\hat{\varphi}_R-i\hat{\varphi}_I,\hat{\varphi}_I\right].
\end{eqnarray}

Though the form of $P$ is not known, and indeed a closed-form expression is unlikely to exist for arbitrary $H$, it is possible to define a Langevin
dynamics of the complex field variables that will sample it, provided it exists.
We begin with a Chapman-Kolmogorov (CK) equation specifying the time evolution of $P$.
Defining a four-component state vector according to $\textbf{x}=(\varphi_R,\varphi_I,\hat{\varphi}_R,\hat{\varphi}_I)^T$, the CK equation can be written
\begin{eqnarray}\label{a5}
& &P\left[\textbf{x},t+\Delta t\right]=\nonumber\\
& &\int \mathcal{D}(\Delta\textbf{x})\Phi\left[\Delta\textbf{x};\textbf{x}-\Delta\textbf{x}\right]
P\left[\textbf{x}-\Delta\textbf{x},t\right],
\end{eqnarray}
where $\Phi\left[\Delta\textbf{x};\textbf{x}\right]$ is the transition
probability density for a displacement $\Delta\textbf{x}$ in the complex plane, starting
at the point $\textbf{x}$, over a time interval of $\Delta t$.
Note that $\textbf{x}$ is a function of $\vect{r},s$ in the present theory.

Eqn.~\ref{a5} can be converted to differential form, a Fokker-Planck equation,
by expanding the left-hand side in powers of $\Delta t$ to $\mathcal{O}\left(\Delta t\right)$, and expanding the right-hand side in
powers of $\Delta \textbf{x}$ to $\mathcal{O}\left(\Delta \textbf{x}^2\right)$. This leads to
\begin{eqnarray}\label{a6}
& &\Delta t\frac{\partial}{\partial t}P\left[\textbf{x},t\right]=-\nabla_{\textbf{x}}\cdot\left[\langle\Delta\textbf{x}\rangle_\Phi P\left[\textbf{x},t\right]\right]\nonumber\\
& &+\frac{1}{2!}\nabla_{\textbf{x}}\nabla_{\textbf{x}}:\left[\langle\Delta\textbf{x}\Delta\textbf{x}\rangle_\Phi P\left[\textbf{x},t\right]\right]+O((\Delta t)^2).
\end{eqnarray}
The first two moments of $\Delta\textbf{x}$ can be obtained from the Langevin dynamics that samples $P$, with averages taken over all
realizations of the random noise.
Our CL dynamics is written
\begin{eqnarray}\label{a7}
\partial_t\varphi(\textbf{r},s,t)&=&-i\frac{\delta H}{\delta\hat{\varphi}}+\mu(\textbf{r},s,t)\nonumber\\
\partial_t\hat{\varphi}(\textbf{r},s,t)&=&-i\frac{\delta H}{\delta\varphi}+\hat{\mu}(\textbf{r},s,t)
\end{eqnarray}
where $\mu(\textbf{r},s,t)$ and $\hat{\mu}(\textbf{r},s,t)$ are complex random variables with
$\mu_1(\textbf{r},s,t)$, $\mu_2(\textbf{r},s,t)$ and $\mu_3(\textbf{r},s,t)$, $\mu_4(\textbf{r},s,t)$
corresponding to their real and imaginary parts, separately.
In our work, each of the four random variables is a white, Gaussian noise with first and second moments
defined as
\begin{eqnarray}\label{a8}
\langle\mu_i(\textbf{r},s,t)\rangle&=&0\nonumber\\
\langle\mu_i(\textbf{r},s,t)\mu_j(\textbf{r}^\prime,s^\prime,t^\prime)\rangle&=&M_{ij}\delta(\textbf{r}-\textbf{r}^\prime)
\delta(s-s^\prime)\delta(t-t^\prime)\nonumber\\
i,j&=&1,2,3,4,
\end{eqnarray}
where $M_{ij}$ is the tensor of correlations between the random variables.
The goal is to define the entries of $M$ that will produce a steady-state distribution of $\textbf{x}$ equal to $P$ (and therefore consistent
with the target $P_C$).
The first moments of $\textbf{x}$ are all related to relaxational terms.
Defining $F_R=\Re\left(-\frac{\delta H}{\delta\varphi}\right)$,
$F_I=\Im\left(-\frac{\delta H}{\delta\varphi}\right)$,
$\hat{F}_R=\Re\left(-\frac{\delta H}{\delta\hat{\varphi}}\right)$ and
$\hat{F}_I=\Im\left(-\frac{\delta H}{\delta\hat{\varphi}}\right)$, the first moments are (to $\mathcal{O}\left(\Delta t\right)$):
\begin{eqnarray}\label{a9}
\langle\Delta\varphi_R\rangle=-\Delta t \hat{F}_I;\langle\Delta\varphi_I\rangle&=&\Delta t \hat{F}_R\\
\langle\Delta\hat{\varphi}_R\rangle=-\Delta t F_I;\langle\Delta\hat{\varphi}_I\rangle&=&\Delta t F_R.
\end{eqnarray}
Likewise, the second moments, taken to  $O(\Delta t)$, are $\left<\Delta \phi_i \Delta \phi_j\right> = M_{ij}\Delta t$.

Inserting the first and second moments of $\textbf{x}$ into Eqn.~\ref{a6} and applying the integral operation of Eqn.\ \ref{a4} to both sides,
we obtain the following Fokker-Planck equation for $P_C$
\begin{eqnarray}\label{a11}
  \frac{\partial}{\partial t}&P&_C\left(\varphi_R,\hat{\varphi}_R,t\right)=i\frac{\partial}{\partial\varphi_R}
\left[\frac{\partial H}{\partial\hat{\varphi}_R}P_C\right]+i\frac{\partial}{\partial\hat{\varphi}_R}
\left[\frac{\partial H}{\partial\varphi_R}P_C\right]\nonumber\\& &+\frac{1}{2}\left(M_{11}-M_{22}+i2M_{12}\right)
\frac{\partial^2}{\partial\varphi^2_R}P_C\nonumber\\
& &+\frac{1}{2}\left(M_{33}-M_{44}+i2M_{34}\right)
\frac{\partial^2}{\partial\hat{\varphi}^2_R}P_C\nonumber\\
& &+\left[M_{13}-M_{24}+i(M_{14}+M_{23})\right]\frac{\partial^2}{\partial\varphi_R\partial\hat{\varphi}_R}P_C\nonumber\\
\end{eqnarray}
In order for Eqn.~\ref{a11} to have a steady state ($\partial_t P_C=0$) solution
$P_C\left(\varphi_R,\hat{\varphi}_R\right)\propto e^{-H(\varphi_R,\hat{\varphi}_R)}$, corresponding to Eqn.~\ref{a2},
the $M$ tensor should therefore satisfy the following conditions
\begin{eqnarray}\label{a12}
  M_{11}-M_{22} &=& 0\\
  M_{33}-M_{44} &=& 0\\
  M_{34}=M_{12} &=& 0\\
  M_{13}-M_{24} &=& 0\\
  M_{14}+M_{23} &=& 1
\end{eqnarray}
so that, noting the symmetry of $M$, we have
\begin{equation}
  M=\left(\begin{array}{cccc}a&0&c&d\\0&a&1-d&c\\c&1-d&b&0\\d&c&0&b\end{array}\right)
\end{equation}
for $a$,$b$,$c$ and $d$ presently unspecified.

Having identified the appropriate noise statistics, we use the following method to generate such correlated random variables from
readily-generated decorrelated random variables (Ref.\ \onlinecite{Franklin1965}).
Let $\eta_i\left(\vect{r},s,t\right)$ be a set of four \emph{uncorrelated} Gaussian-distributed random variables with moments
$\left<\eta_i\right>=0$, $\left<\eta_i\left(\vect{r},s,t\right)\eta_j\left(\vect{r}^\prime,s^\prime,t^\prime\right)\right> = 2\delta_{i,j}\delta\left(\vect{r}-\vect{r}^\prime\right)\delta\left(s-s^\prime\right)\delta\left(t-t^\prime\right)$.
We are required to transform to the correlated noise set, $\mu_i = \sum_j L_{ij} \eta_j$, so that
$\left<\mu_i \mu_j\right> = \left<\sum_{k}\sum_{l} L_{ik}L_{jl}\eta_k\eta_l\right> = \sum_k L_{ik} L_{jk} = M_{ij}$
The latter can be obtained by symmetric factorization of $M$ through Cholesky decomposition ($M=LL^T$) if $M$ is Hermitian and positive semi-definite.
This places restrictions on the values of $a$,$b$,$c$ and $d$ that may be chosen.
In our investigations, we have found the choice $a=b=d=1/2$ and $c=0$ to be most efficient.
This yields
\begin{eqnarray}\label{a17}
M=\frac{1}{2}\left(\begin{array}{cccc}
1&0&0&1\\
0&1&1&0\\
0&1&1&0\\
1&0&0&1
\end{array}
\right)
\end{eqnarray}
and
\begin{eqnarray}\label{a18}
L=\frac{1}{\sqrt{2}}\left(\begin{array}{cccc}
1&0&0&0\\
0&1&0&0\\
0&1&0&0\\
1&0&0&0
\end{array}
\right)
\end{eqnarray}
Notice that the decomposed $L$ matrix has two columns consisting of zeros, meaning that
$\eta_3$ and $\eta_4$ need not be applied during the CL dynamics.
Hence, this particular choice of $L$ leads to less noise application, and consequently
lower population variance, than the general case, while maintaining thermodynamic consistency.
Finally, we can summarize the appied noise $\vect{\mu}$ which satisfies the conditions thermodynamic consistency
\begin{eqnarray}\label{a19}
\vect{\mu}={\text{L}}{\eta}=\frac{1}{\sqrt{2}}\left(\begin{array}{c}
\eta_1\\
\eta_2\\
\eta_2\\
\eta_1
\end{array}
\right)
\end{eqnarray}
as found in Eqn.\ \ref{f25}.

\section{Numerical Discretization of CS-CL Dynamics}
\label{sec:discrete}
In this appendix, we detail the numerical scheme used in the present paper for solving the
compound CL dynamics with stabilizing non-Hermitian mobility matrix included.
The correct discretized distributions of the Langevin noise terms, with continuum limit
consistent with Eqn.~\ref{f25}, are motivated here.
For generality, we proceed with explicit inclusion of the forcing source terms of the action (i.e., the ``free boundary'' method).
The continuum CS-CL dynamics equations that we aim to solve are,
\begin{eqnarray}
  \partial_t\varphi\left(\vect{r},s,t\right)&=&-\left(\partial_s-\nabla^2+w(\vect{r})\right)\varphi\left(\vect{r},s,t\right)\nonumber\\
                         &+&\sqrt{z}\delta(s)+\mu(\vect{r},s,t)\\
  \partial_t\hat{\varphi}\left(\vect{r},s,t\right)&=&\left(\partial_s+\nabla^2-w(\vect{r})\right)\hat{\varphi}\left(\vect{r},s,t\right)\nonumber\\
                             &+&i\sqrt{z}\delta(s-1)+\hat{\mu}(\vect{r},s,t),
\end{eqnarray}
where $w(\vect{r})=B\Gamma(\vect{r})\ast\hat{\rho}(\vect{r})$, and $\mu\left(\vect{r},s,t\right)$ and $\hat{\mu}\left(\vect{r},s,t\right)$
are two complex-valued random variables defined in Eqn.~\ref{f25}.

A first-order semi-implicit forward-Euler integration method\cite{Lennon2008b} was used to time step the discretized, stochastic CL equations.
We consider in detail only the equation of motion for $\varphi$. For improved stability, we
add to the linear part of the force at the future time the maximum value of $w(\vect{r})$, $w_\mathrm{max}$,
and subtract the same term at the present time.
The field update expression becomes
%
\begin{eqnarray}
\left(1+\Delta t\left(\partial_s -\nabla^2 + w^t_\mathrm{max}\right)\right)\varphi^{\vect{r},t+\Delta t}(s) =\nonumber\\
\left(1-\Delta t \delta w^{\vect{r},t}\right)\varphi^{\vect{r},t}\left(s\right)+\Delta t \sqrt{z}\delta\left(s\right) + \mu^{\vect{r},t}\left(s\right),
\end{eqnarray}
with
\begin{eqnarray}
\left<\mu^{\vect{r},t}\left(s\right)\right>&=&0\nonumber\\
  \left<\Re\left(\mu^{\vect{r},t}\left(s\right)\right)\Re\left(\mu^{\vect{r^{\prime}},t^{\prime}}\left(s^{\prime}\right)\right)\right>&=&\frac{\Delta t}{\Delta V}\delta\left(s-s^{\prime}\right)\delta_{\vect{r},\vect{r^{\prime}}}\delta_{t,t^{\prime}},\nonumber\\
  \\
  \left<\Im\left(\mu^{\vect{r},t}\left(s\right)\right)\Im\left(\mu^{\vect{r^{\prime}},t^{\prime}}\left(s^{\prime}\right)\right)\right>&=&\frac{\Delta t}{\Delta V}\delta\left(s-s^{\prime}\right)\delta_{\vect{r},\vect{r^{\prime}}}\delta_{t,t^{\prime}},\nonumber\\
  \\
  \left<\Im\left(\mu^{\vect{r},t}\left(s\right)\right)\Re\left(\mu^{\vect{r^{\prime}},t^{\prime}}\left(s^{\prime}\right)\right)\right>&=&0,
\end{eqnarray}
where $\delta w^{\vect{r},t}=w^{\vect{r},t}-w^t_\mathrm{max}$, $\Delta t$ is the time step and $\Delta V$ is
the volume element associated with a sample on the spatial collocation mesh.
Notice that the second moment of the applied noise does not have the factor $2$, which is cancelled by the $1/\sqrt{2}$ found in the correlation constraints
(Eqn.\ \ref{a19}).
We note that the linear-order force terms are diagonal in Fourier space, and
hence it is most convenient to consider the Fourier-transformed CS-CL equations
\begin{eqnarray}\label{f29}
\left(1+\Delta t\left(\partial_s +\vect{k}^2 + w^t_\mathrm{max}\right)\right)\varphi^{\vect{k},t+\Delta t}(s) =\nonumber\\
\mathcal{F}\left[\left(1-\Delta t \delta w^{\vect{r},t}\right)\varphi^{\vect{r},t}\left(s\right) + \mu^{\vect{r},t}\left(s\right)\right]
+\Delta t \sqrt{z}\delta\left(s\right)\delta_{\vect{k},0},\nonumber\\
\end{eqnarray}
where $\mathcal{F}$ is the discrete Fourier transform.
Our discrete Fourier transform conventions are
\begin{eqnarray}
  f_\vect{r} = \sum_\vect{k}\tilde{f}_\vect{k} e^{i\vect{k}.\vect{r}} \\
  \tilde{f}_\vect{k} = \frac{1}{M}\sum_\vect{r}f_\vect{r} e^{-i\vect{k}.\vect{r}},
\end{eqnarray}
with $\vect{k}$ in the set $\frac{2\pi}{L}\left(i\hat{x}+j\hat{y}+k\hat{z}\right)$ for our cubic simulation cell,
with integers $i,j,k \in \left[ -N_x/2, N_x/2\right]$, unit vectors in reciprocal space, $\left\{\hat{x}, \hat{y}, \hat{z}\right\}$,
and $M=N_x^3$, $V=L^3$.
In the present implementation we use the \verb!FFTW! library to execute these transforms\cite{FrigoJohnson2005}.
With these conventions, the $\vect{k}=0$ mode of any field corresponds directly to the
volume-normalized spatial average in $\vect{r}$.

To discretize the contour variable, we use a first-order forward-Euler stepping scheme.
Taking Eqn.~\ref{f29} and rearranging
\begin{eqnarray}\label{f30}
\partial_s \varphi^{\vect{k},t+\Delta t}(s)=
-\frac{1}{\Delta t}\left(1+\Delta t\left(\vect{k}^2 + w^t_\mathrm{max}\right)\right)\varphi^{\vect{k},t+\Delta t}\left(s\right)\nonumber\\
+\frac{1}{\Delta t}\mathcal{F}\left[\left(1-\Delta t \delta w^{\vect{r},t}\right)\varphi^{\vect{r},t}\left(s\right) + \mu^{\vect{r},t}\left(s\right)\right]\nonumber\\
+\sqrt{z}\delta\left(s\right)\delta_{\vect{k},0}.\nonumber\\
\end{eqnarray}
An implicit forward-Euler scheme for $s$ is
\begin{eqnarray}\label{f31}
\frac{\varphi^{\vect{k},t+\Delta t}_s-\varphi^{\vect{k},t+\Delta t}_{s-1}}{\Delta s}=
-\frac{1}{\Delta t}\left(1+\Delta t\left(\vect{k}^2 + w^t_\mathrm{max}\right)\right)\varphi^{\vect{k},t+\Delta t}_s\nonumber\\
+\frac{1}{\Delta t}\mathcal{F}\left[\left(1-\Delta t \delta w^{\vect{r},t}\right)\varphi^{\vect{r},t}_s + \mu^{\vect{r},t}_s\right]\nonumber\\
+\frac{\sqrt{z}}{\Delta s}\delta_{s,0}\delta_{\vect{k},0}\nonumber\\
\end{eqnarray}
with
\begin{eqnarray}\label{f32}
\left<\mu^{\vect{r},t}_s\right>&=&0\\
  \left<\Re\left(\mu^{\vect{r},t}_s\right)\Re\left(\mu^{\vect{r^{\prime}},t^{\prime}}_{s^{\prime}}\right)\right>&=&\frac{\Delta t}{\Delta s\Delta V}\delta_{s,s^{\prime}}\delta_{\vect{r},\vect{r^{\prime}}}\delta_{t,t^{\prime}}\\
  \left<\Im\left(\mu^{\vect{r},t}_s\right)\Im\left(\mu^{\vect{r^{\prime}},t^{\prime}}_{s^{\prime}}\right)\right>&=&\frac{\Delta t}{\Delta s\Delta V}\delta_{s,s^{\prime}}\delta_{\vect{r},\vect{r^{\prime}}}\delta_{t,t^{\prime}}\\
  \left<\Re\left(\mu^{\vect{r},t}_s\right)\Im\left(\mu^{\vect{r^{\prime}},t^{\prime}}_{s^{\prime}}\right)\right>&=&0,
\end{eqnarray}
where $\Delta s$ is the contour variable step size.
Then the final update scheme for $\varphi^{\vect{k},t+\Delta t}_s$ is
\begin{eqnarray}\label{f33}
  \varphi^{\vect{k},t+\Delta t}_s = \frac{v^{\vect{k},t}_{s} + \frac{\Delta t}{\Delta s}\varphi^{\vect{k},t+\Delta t}_{s-1}}{d^{\vect{k},t}}
\end{eqnarray}
where
\begin{eqnarray}\label{f34}
  d^{\vect{k},t}= \left(1+\Delta t\left(\vect{k}^2 + w^t_\mathrm{max}\right) + \frac{\Delta t}{\Delta s}\right)
\end{eqnarray}
and
\begin{eqnarray}\label{f35}
  v^{\vect{k},t}_{s} =\mathcal{F}\left[\left(1-\Delta t\delta w^{\vect{r},t}\right)\varphi^{\vect{r},t}_s  + \mu^{\vect{r},t}_s\right]
  + \frac{\Delta t}{\Delta s}\sqrt{z}\delta_{\vect{k},0}\delta_{s,0}\nonumber\\
\end{eqnarray}

Note that this expression is entirely local in $\vect{k}$. Everything on the RHS is either evaluated at the
previous CL time step $t$, or at the future time step $t+\Delta t$ for an earlier contour position.
Since the update scheme depends on $s-\Delta s$, we must consider an initial condition for $s$.
In the present method, we set $\varphi\left(\vect{r},s\right) = 0 \quad \forall \, s<0$ (i.e., for contour positions outside the domain of physical interest, and
causally earlier than the source-injecting $\sqrt{z}$ term.

The discretized equation of motion for $\hat{\varphi}$ is developed in a similar way, with the only difference that we use a backward-difference Euler stepping
scheme in the contour variable.
This choice results in the target causal property for $G$ (i.e., $G=0$ for $s\leq s^\prime$,
specifically with $0$-valued diagonal elements).
A final technical consideration is on the correlated noise.
Eqn.\ \ref{f25} shows that $\varphi$ and $\hat{\varphi}$ have correlated noise applied at equal $s$.
Since we have a forward-stepping scheme for disctretizing $s$ in $\varphi$, and a backward-stepping scheme for
discretizing $s$ in $\hat{\varphi}$, it is required for the real (imaginary) part of the
noise applied at the $\varphi_s$ grid point to be applied
to the imaginary (real) part of the $\hat{\varphi}_{s-1}$ grid point.

For the ``fixed boundary'' method, we simply eliminate the noise term applied to
the $\varphi_{s=0}$ and $\hat{\varphi}_{s=N_s}$ modes.
This choice causes those modes to quickly relax during thermalization
to a static, spatially homogeneous field value that is consistent with the source term
integrated over the contour $\delta$ functions,
with field values strictly pinned to $0$ at causally earlier contour positions.


\end{document}